\documentclass[apjl]{emulateapj}

\slugcomment{To appear in ApJ Letters}

\usepackage{graphicx}
\usepackage{apjfonts}
\usepackage{amsmath}
\usepackage[raggedright]{subfigure}
\usepackage{color}

\newcommand{\psr}{PSR~J1838$-$0537}
\newcommand{\fgl}{2FGL~J1839.0$-$0539}
\newcommand{\tev}{HESS~J1841$-$055}

\def\EAH{{\it Einstein@Home}}
\def\emax{E_{\rm max}}

\newcommand{\glf}{5.5 \times 10^{-6}}
\newcommand{\glfdot}{1.5\%}
\newcommand{\glfddot}{2.4}

\shorttitle{\psr} 
\shortauthors{\sc Pletsch et al.}

\begin{document}

\title{\psr: Discovery of a young, energetic gamma-ray pulsar
} 

\author{
H.~J.~Pletsch\altaffilmark{1,2,3},
L.~Guillemot\altaffilmark{4,5},
B.~Allen\altaffilmark{1,6,2},
M.~Kramer\altaffilmark{4,7},
C.~Aulbert\altaffilmark{1,2},
H.~Fehrmann\altaffilmark{1,2},
M.~G.~Baring\altaffilmark{8},
F.~Camilo\altaffilmark{9},
P.~A.~Caraveo\altaffilmark{10},
J.~E.~Grove\altaffilmark{11},
M.~Kerr\altaffilmark{12},
M.~Marelli\altaffilmark{10},
S.~M.~Ransom\altaffilmark{13},
P.~S.~Ray\altaffilmark{11}, and
P.~M.~Saz~Parkinson\altaffilmark{14}
}
\altaffiltext{1}{Max-Planck-Institut f\"ur Gravitationsphysik (Albert-Einstein-Institut), 30167 Hannover, Germany}
\altaffiltext{2}{Leibniz Universit\"at Hannover, 30167 Hannover, Germany}
\altaffiltext{3}{email: holger.pletsch@aei.mpg.de}
\altaffiltext{4}{Max-Planck-Institut f\"ur Radioastronomie, Auf dem H\"ugel 69, 53121 Bonn, Germany}
\altaffiltext{5}{email: guillemo@mpifr-bonn.mpg.de}
\altaffiltext{6}{Department of Physics, University of Wisconsin-Milwaukee, P.O. Box 413, Milwaukee, WI 53201, USA}
\altaffiltext{7}{Jodrell Bank Centre for Astrophysics, School of Physics and Astronomy, The University of Manchester, M13 9PL, UK}
\altaffiltext{8}{Rice University, Department of Physics and Astronomy, MS-108, P. O. Box 1892, Houston, TX 77251, USA}
\altaffiltext{9}{Columbia Astrophysics Laboratory, Columbia University, New York, NY 10027, USA}
\altaffiltext{10}{INAF-Istituto di Astrofisica Spaziale e Fisica Cosmica, 20133 Milano, Italy}
\altaffiltext{11}{Space Science Division, Naval Research Laboratory, Washington, DC 20375-5352, USA}
\altaffiltext{12}{W. W. Hansen Experimental Physics Laboratory, Kavli Institute for Particle Astrophysics and Cosmology, Department of Physics and SLAC National Accelerator Laboratory, Stanford University, Stanford, CA 94305, USA}
\altaffiltext{13}{National Radio Astronomy Observatory (NRAO), Charlottesville, VA 22903, USA}
\altaffiltext{14}{Santa Cruz Institute for Particle Physics, Department of Physics and Department of Astronomy and Astrophysics, University of California at Santa Cruz, Santa Cruz, CA 95064, USA}

\begin{abstract} 
\noindent
We report the discovery of \psr, a gamma-ray pulsar found through a blind search 
of data from the \textit{Fermi} Large Area Telescope~(LAT).
The pulsar has a spin frequency of  6.9\,Hz and a frequency derivative of 
$-2.2\times 10^{-11}$\,Hz s$^{-1}$, implying a young characteristic age of 4970~years
and a large spin-down power of \mbox{$5.9\times10^{36}$\,erg s$^{-1}$}.
Follow-up observations with radio telescopes detected no pulsations,
thus \psr{} appears radio-quiet as viewed from Earth.
In September 2009 the pulsar suffered the largest glitch so far seen in any 
gamma-ray-only pulsar, causing a relative increase in spin frequency of about $\glf$.
After the glitch, during a putative recovery period, the timing analysis 
is complicated by the sparsity of the LAT photon data,
the weakness of the pulsations, and the reduction in average exposure 
from a coincidental, contemporaneous change in the LAT's sky-survey observing pattern.
The pulsar's sky position is coincident with the spatially extended 
TeV source \tev{} detected by the High Energy Stereoscopic System (H.E.S.S.).
The inferred energetics suggest that \tev{} contains
a pulsar wind nebula powered by the pulsar.
\end{abstract}


\section{Introduction}\label{s:intro}

In the last three years, the Large Area Telescope \citep[LAT;][]{generalfermilatref} 
aboard the {\em Fermi} satellite has proven a revolutionary detector of 
gamma-ray pulsars\footnote{See https://confluence.slac.stanford.edu/display/GLAMCOG/Public+ List+of+LAT-Detected+Gamma-Ray+Pulsars/}.
These objects are rapidly rotating, highly magnetized neutron stars,
whose rotation carries the gamma-ray emitting regions past 
an observer's line of sight, creating periodic pulsations. 

With the LAT, pulsars have been detected via three different approaches.  
First, for pulsars known from radio observations,
gamma-ray pulsations are detected by coherent folding of the 
LAT photon arrival times based on the provided pulsar ephemerides
\citep[e.g.,][]{FermiPSRCatalog,Guillemot2012}. 
Second, radio searches of unassociated gamma-ray  sources, 
as listed in the {\em Fermi}-LAT Second 
Source Catalog \citep[2FGL;][]{FermiSecondSourceCatalog}, have
revealed new radio pulsars. The ephemerides obtained from radio 
timing observations are in turn used to coherently phase-fold 
the LAT data and probe for gamma-ray pulsations
\citep[e.g.;][]{Ransom2011,Cognard2011,Camilo2012,Kerr2012,Ray+2012}.

Contrasting the previous two strategies, ``blind searches'' are not guided 
by prior knowledge of the pulsar parameters \citep[see e.g.][]{Chandler2001}.
Within a year after launch of {\em Fermi}, 
previous blind searches had impressively unveiled
24 pulsars \citep[][]{16gammapuls2009,8gammapuls2010} 
using a time-differencing technique \citep{Atwood2006,Ziegler2008}. 
Another two pulsars were found after 
two years \citep{2gammapuls2012}, however the detection rate had dropped since then, 
predominantly because of the increasing computational challenge with the longer data time span.

The sensitivity of blind searches for gamma-ray pulsars is limited by
computing cost, because a grid in the search parameter space 
(for isolated systems typically sky location, frequency $f$, 
and spin-down rate $\dot f$) must be explicitly searched. 
The number of grid points required to discretely cover the relevant 
parameter space increases as a high power of the coherent integration 
time \citep[e.g.,][]{bccs1:1998}. 
This problem is analogous to blind searches for continuous gravitational waves (CWs)
emitted from rapidly spinning neutron stars,
where hierarchical strategies \citep{schutzpapa:1999,bc2:2000,Houghpaper,cutler:2005} 
and semi-coherent methods are used to efficiently scan wide parameter-space ranges
in years of data at finite computational resources.

Here, we present the discovery and key parameters of a young, 
energetic gamma-ray pulsar, \psr, detected during a new blind survey 
exploiting a novel data-analysis method \citep{Pletsch9pulsars}
originally developed to blindly search for weak CW signals \citep{PletschAllen2009,Pletsch2010,PletschSLCW2011}.
The pulsar is spatially coincident with the extended source \tev{} 
\citep{Aharonian2008HESS}, detected at very high energies (VHE; $E >0.1$\,TeV). 
We show that the pulsar's energetics make this association plausible.

\section{Discovery of \psr}\label{s:disco}

In recent work \citep{Pletsch9pulsars} we reported on the discovery of 
nine new gamma-ray pulsars in this novel, ongoing blind survey 
using 975 days of LAT data. Unlike those nine pulsars, \psr{},
being significantly younger, was initially detected only during the
first year of data (before its large glitch). 
After the glitch, recovering the (weaker) pulsations proved to be a substantial  
challenge requiring a dedicated follow-up study presented below 
(Section~\ref{s:fu-analysis}).

In the new blind-search effort, detecting \psr{},
we have utilized a hierarchical (multistage) approach.
The first stage is semi-coherent, where coherent Fourier powers 
computed over a 6-day window are incoherently combined 
by sliding the window over the entire 975 days of data. 
A fundamental new element of the search method is the exploitation of
a parameter-space metric  \citep{PletschAllen2009,Pletsch2010} to build 
an efficient search grid (in $f$, $\dot f$ \emph{and} sky position).
In a second stage, significant semi-coherent candidates are automatically 
followed up via a fully coherent analysis. A third stage further refines coherent 
pulsar candidates by using higher signal harmonics adopting the  
$H$-test \citep{deJaeger1989}. Further advances incorporated are
described in \citet{Pletsch9pulsars}.

The survey targets unassociated sources from the 2FGL catalog 
with ``pulsar-like'' properties. Such sources feature 
significantly curved emission spectra and low flux variability over time. 
Further details on the survey regarding 
source selection, data preparation and search parameter-space 
are found in \citet{Pletsch9pulsars}.

We have detected \psr{} as part of the survey in the blind search of \fgl. 
This source also has gamma-ray counterparts 
1FGL J1839.1$-$0543c in the \emph{Fermi}-LAT First Source Catalog 
\citep[][]{FermiFirstSourceCatalog} and 0FGL J1839.0$-$0549 
in the \emph{Fermi} Bright Source List \citep[][]{FermiBSL}.

\begin{figure*}
\centerline{
\hfill
\includegraphics[width=0.93\textwidth]{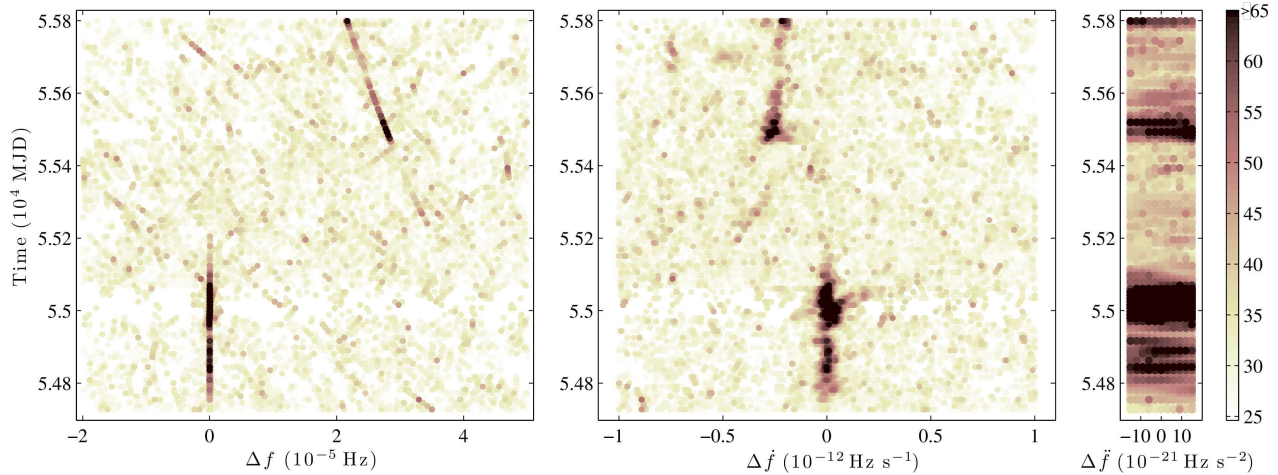}
\hfill
}
\caption{\label{f:glitch} 
Pulsar glitch analysis. 
Color-coded is the weighted $H$-test statistic for photons
lying within a 90-day window that is slid over the entire data set with 90\% overlap. 
Fixing the sky position at the pre-glitch (before MJD 55100)
solution, for each window scans in $H$-test over $\{f,\dot f,\ddot f\}$ are done.  
Vertical axes show the time midpoint of each window. 
Horizontal axes show the offsets from the inferred pre-glitch
parameters in $f$ (left), $\dot f$ (middle), and $\ddot f$ (right).\\
}
\end{figure*}

\begin{deluxetable}{ll}
\tablewidth{\columnwidth}
\tablecaption{\label{t:parameters} Parameters of \psr }
\tablecolumns{2}
\tablehead{
\colhead{Parameter} &
\colhead{Value}
}
\startdata
Right ascension, $\alpha$ (J2000.0)\dotfill & $18^{\rm h}38^{\rm m}56\fs02(3)$\\
Declination, $\delta$ (J2000.0)\dotfill    & $-05\arcdeg37\arcmin09\arcsec(2)$\\
Galactic longitude, $l$ (\arcdeg)\dotfill   & 26.5 \\
Galactic latitude, $b$ (\arcdeg)\dotfill    & 0.2 \\
Spin frequency, $f$ (Hz)\dotfill 
 & $6.863015715(4)$\tablenotemark{a} \\
& $6.86305339(1)$\tablenotemark{b}\\
Frequency 1st derivative, $\dot f$ ($10^{-11}$ Hz $s^{-1}$)\dotfill 
& $-2.18964(6)$\tablenotemark{a}  \\
& $-2.2222(1)$\tablenotemark{b}  \\
Frequency 2nd derivative, $\ddot f$ ($10^{-22}$ Hz $s^{-2}$)\dotfill 
& $5.0(4)$\tablenotemark{a}  \\
& $17.0(9)$\tablenotemark{b} \\
Epoch (MJD)\dotfill  & 55100.0     \\
Data span (MJD)\dotfill  & 54702 -- 55836\\
Characteristic age, $\tau_c$ (yr)\dotfill  & $4970$      \\
Spin-down power, $\dot E$ (${\rm erg\,s^{-1}}$)\dotfill & $5.9\times10^{36}$ \\
Surface magnetic field strength, $B_{\textrm{S}}$ (G)\dotfill & $8.3\times10^{12}$ \\
Light-cylinder magnetic field strength, $B_{\textrm{LC}}$ (kG)\dotfill & $24.7$ \\
Photon index, $\Gamma$ \dotfill & $1.8 \pm 0.1$ \\
Cutoff energy, $E_c$ (GeV) \dotfill  & $5.6 \pm 1.4 $  \\
Photon flux above 100\,MeV ($10^{-8}$ photons cm$^{-2}$ s$^{-1}$)\dotfill & $23.9 \pm 5.1$ \\
Energy flux above 100\,MeV ($10^{-11}$ erg cm$^{-2}$ s$^{-1}$) \dotfill & $17.5 \pm 1.8$\\
\enddata
\tablecomments{Numbers in parentheses are statistical 1$\sigma$ errors in the last digits.}
\tablenotetext{a}{Pre-glitch solution (before MJD 55100).}
\tablenotetext{b}{Post-glitch solution (after MJD 55450).\\
}
\end{deluxetable}

\section{Follow-Up Analysis}\label{s:fu-analysis}

\subsection{Data Preparation}
\label{s:dataprep}

For the follow-up study, we prepared a dedicated LAT data set
extending the blind-search input data to 1168~days, up to 2011 October~17.
Using the \emph{Fermi} 
Science Tools\footnote{http://fermi.gsfc.nasa.gov/ssc/data/analysis/scitools/overview.html},
we selected LAT photons whose reconstructed directions 
lie within 15$^\circ$ of the pulsar, have energies above 100~MeV, 
and zenith angles $\leq$ 100$^\circ$. 
We included only ``Source''-class photons according 
to the P7\_V6 Instrument Response Functions and excluded
times when the satellite's rocking angle exceeded 52$^\circ$.

To improve the signal-to-noise ratio, each photon is assigned 
a weight measuring the probability that it has originated from the pulsar \citep{KerrWeightedH2011}.
These weights are computed through a binned likelihood analysis 
(as done for the original blind-search input data) using
a spectral model for the region including 
all 2FGL-catalog sources found within 15$^\circ$ of \psr.
The pulsar spectrum is modeled as an exponentially cutoff power law, 
\mbox{$dN/dE \propto E^{-\Gamma} \exp \left( - E / E_c \right)$}, 
where $\Gamma$ is the photon index and $E_c$ is the cutoff energy. 
The extragalactic diffuse emission and the residual background 
are modeled jointly using the {\em iso\_p7v6source} template, and the Galactic 
diffuse emission is modeled using the {\em gal\_2yearp7v6\_v0} 
map cube. For computational feasibility, only the 165\,000 photons with the 
highest weights are considered (implying a probability-weight threshold of 0.018).

\subsection{Pulsar Timing and Glitch Analysis}

Using the 1168-day data set, the initial pulsar parameters 
are further refined through a timing-analysis procedure.
With techniques described in \citet{Ray2011}, pulse Times Of Arrival (TOAs) 
are precisely measured and parameters of a timing model are fit to these 
measurements.

TOAs are obtained from subdividing the follow-up data set into 
40~segments of about equal length. 
The initial pulsar parameters are used to fold the photon arrival times
and produce a set of pulse profiles. 
By cross-correlating each pulse profile with a multi-Gaussian 
template derived from fitting the entire data set, the TOAs are determined
using the unbinned-maximum-likelihood method of \citet{Ray2011}.  
Then \textsc{Tempo2}~\citep{Tempo2} is used to fit the TOAs to a 
timing model including sky position, frequency and frequency derivatives.

As is typical for a youthful pulsar, \psr{} shows irregularities 
in its spin-frequency evolution that are of two types:
(i) timing noise, observed as slowly varying, non-deterministic fluctuations 
in frequency, which can be modeled by including a second frequency derivative 
in fitting the TOAs; (ii) glitch activity, an abrupt spin-frequency increase, 
typically followed by a recovery phase towards the pre-glitch rotational state. 
Thus, the glitch of \psr{} leads to rapid loss of phase coherence 
in the timing analysis if not additionally accounted for. 

The spin-parameter changes after the glitch are estimated
in a dedicated analysis. 
Fixing the sky position at the pre-glitch (before MJD~55100) 
timing solution, ranges in frequency $f$, frequency derivative $\dot f$, 
and second frequency derivative $\ddot f$ are scanned on a dense grid around their 
pre-glitch values, computing the weighted $H$-test \citep{KerrWeightedH2011}
at each grid point using photons within a fixed time window.
As a balance between signal-to-noise ratio and time resolution, 
we use a window size of 90~days, which is just
long enough to still produce a detectable signal-to-noise ratio.
This 90-day window is slid over the entire data set with 90\% overlap
between subsequent steps. 
The results (Figure~\ref{f:glitch}) indicate that the pulsar 
experienced a glitch near MJD~55100. 
However, immediately after the nominal glitch epoch no convincing signal
is detected for a period of about 300~days. This might be due to
the pulsar's frequency changing faster during early glitch recovery
than the $\{f,\dot f, \ddot f\}$ phase model used. In this respect,
shorter window sizes would possibly help, but these are prohibited
by the sparse sampling of LAT photons.

From Figure~\ref{f:glitch}, estimated values for the glitch parameters are 
added to the timing model, serving as input to iterate 
the above timing procedure. 
Table~\ref{t:parameters} shows the obtained phase-coherent timing solution.
The measured post-glitch spin parameters
imply a relative increase in~$f$ of \mbox{$\Delta f/f \sim \glf$}
and a relative decrease in $\dot f$ of $-\Delta \dot f / \dot f \sim \glfdot$.
In addition, the timing solution requires a relative increase in~$\ddot f$
of $\Delta \ddot f / \ddot f \sim \glfddot$, which likely
accounts mainly for timing noise. In terms of $\Delta f / f$, this glitch is in the 
top 5\% of all pulsar glitches recorded to date \citep{Espinoza2011}, 
and is the largest observed for any gamma-ray-only pulsar \citep[][]{Dormody2012}. 
Measuring such strong pulsar glitches is important, as they may allow probing
the physics of neutron-star interiors \citep[e.g.,][]{ShapiroTeukolsky1983,Haskell2012}.

\begin{figure*}
\centering
	\hspace{-0.2cm}
		\subfigure
		{\includegraphics[width=0.75\columnwidth]{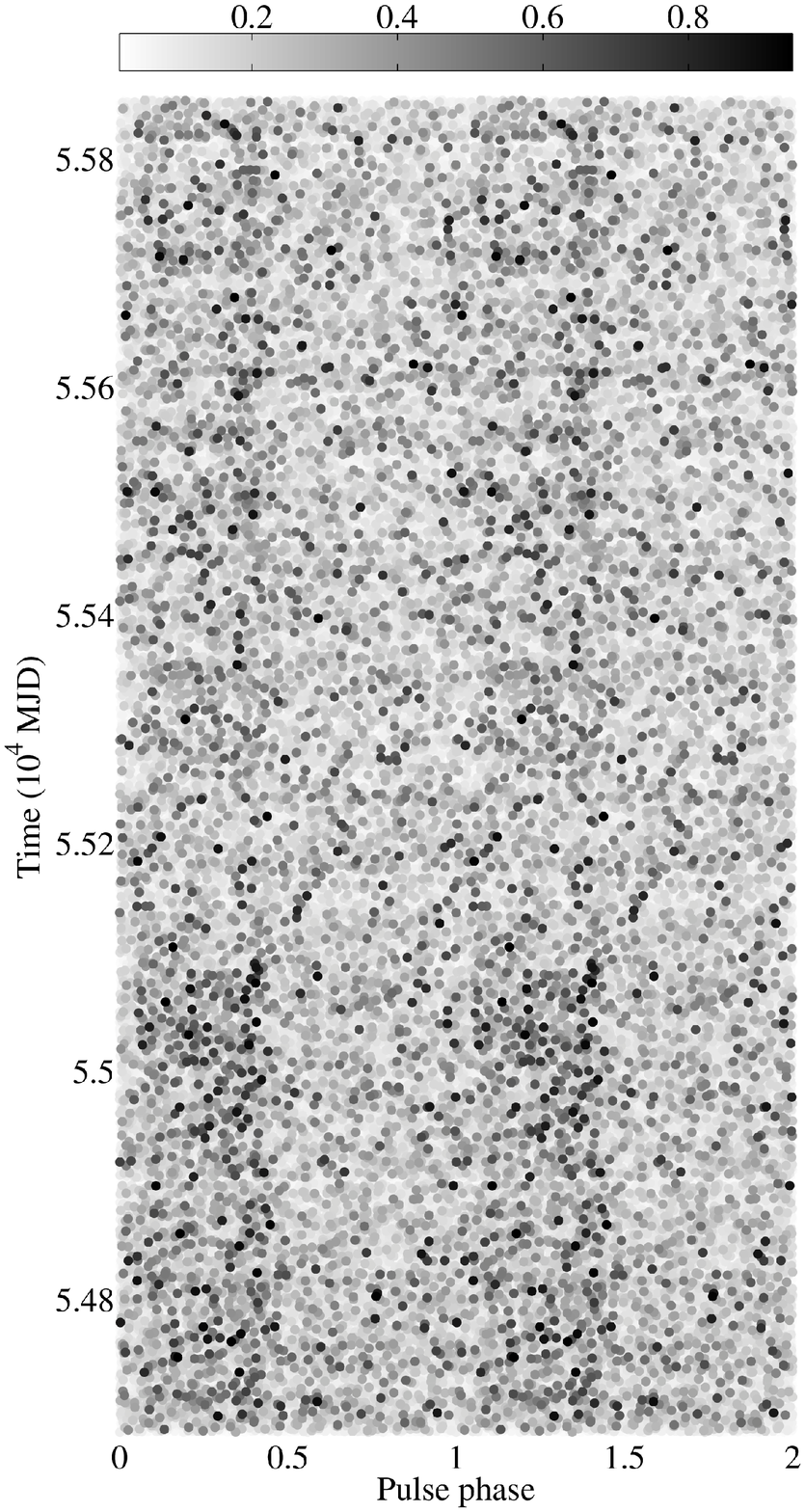}}\hspace{1cm}
		\subfigure
		{\includegraphics[width=0.75\columnwidth]{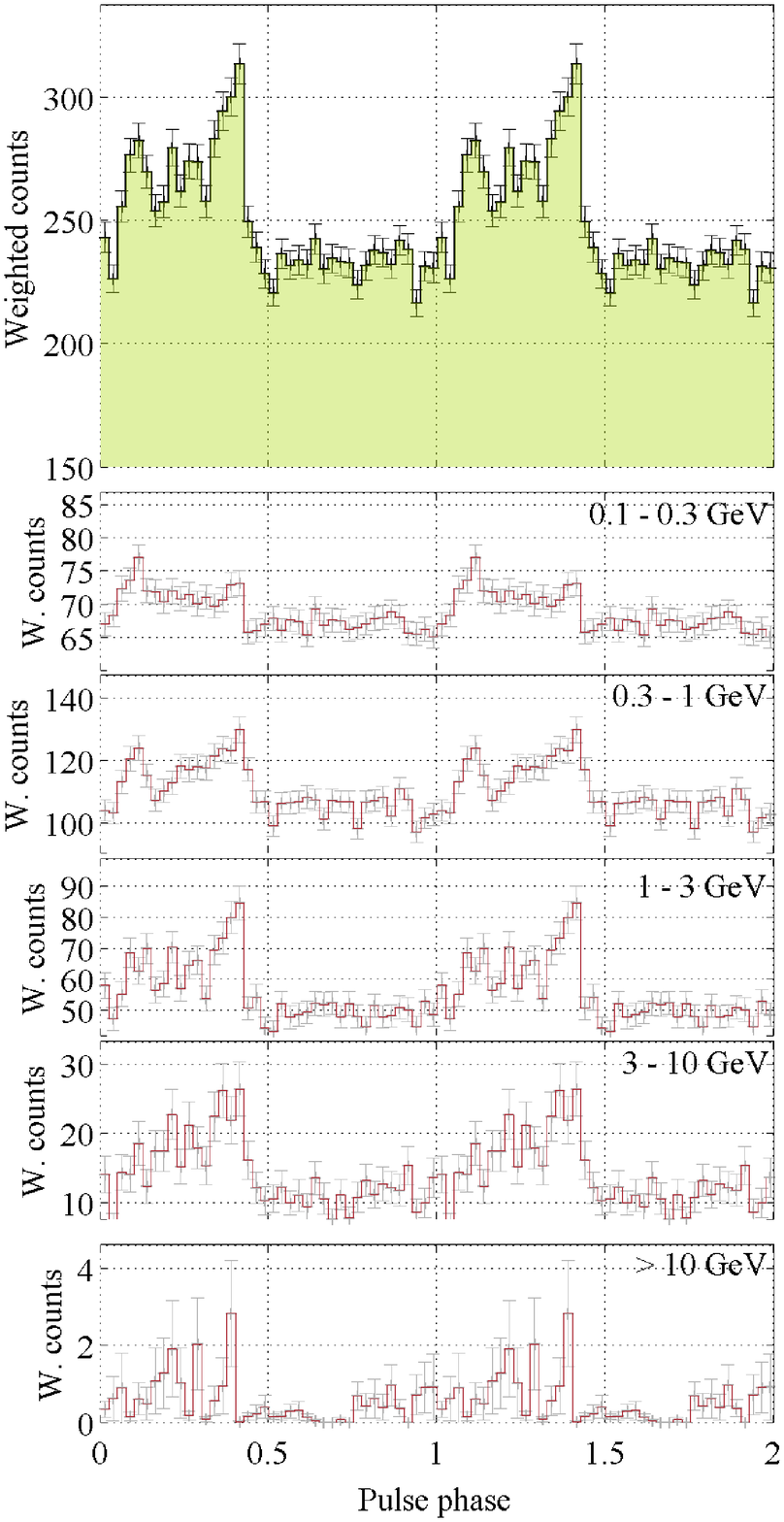}}
	\caption{\label{f:ph-vs-t-all} 
	 Left: Phase-time diagram, showing the pulse phase for each photon arrival time with
	 the photon probability weight represented by gray-scale intensity.
	 Right: The upper plot shows the integrated pulse profile (40 bins per rotation);
          error bars represent 1$\sigma$ statistical uncertainties.
          The five plots below show integrated pulse profiles in different energy ranges. 
          For clarity, horizontal axes always show two rotations.\\
	}
\end{figure*}

The measured parameters listed in Table~\ref{t:parameters} 
characterize the pulsar as young and energetic. 
The implied characteristic age and surface magnetic field strength
are $\tau =-f/2\dot f = 4970\,\textrm{yr}$ and 
$B_{\textrm{S}} = 3.2\times10^{19} (-\dot{f}/f^3 \,\textrm{s}^{-1})^{1/2}\,\textrm{G} = 8.3\times 10^{12}\,\textrm{G}$. Assuming a neutron-star moment of inertia of $I=10^{45}$~g~cm$^2$, 
the pulsar's spin-down power is derived as
 $\dot E = -4\pi^2 I f \dot{f} = 5.9\times 10^{36}$~erg s$^{-1}$.

\subsection{Pulse Profile and Spectral Parameters}

Figure~\ref{f:ph-vs-t-all} shows the phase-time diagram and pulse 
profile using the timing solution of Table~\ref{t:parameters}.
In the integrated pulse profile (weighted pulse phase histogram) statistical 
errors are obtained as $(\sum_j w_j^2)^{1/2}$, where $j$ runs over all photons 
in the same phase bin and $w_j$ denotes the $j$th photon's probability weight.
Comparing the gamma-ray emission before and after the glitch,
we found no observable changes in flux, pulse profile or spectrum of the pulsar,
consistent with other LAT-pulsar glitches \citep[][]{Dormody2012}.

\begin{figure}
\centerline{
\hfill
\includegraphics[width=\columnwidth]{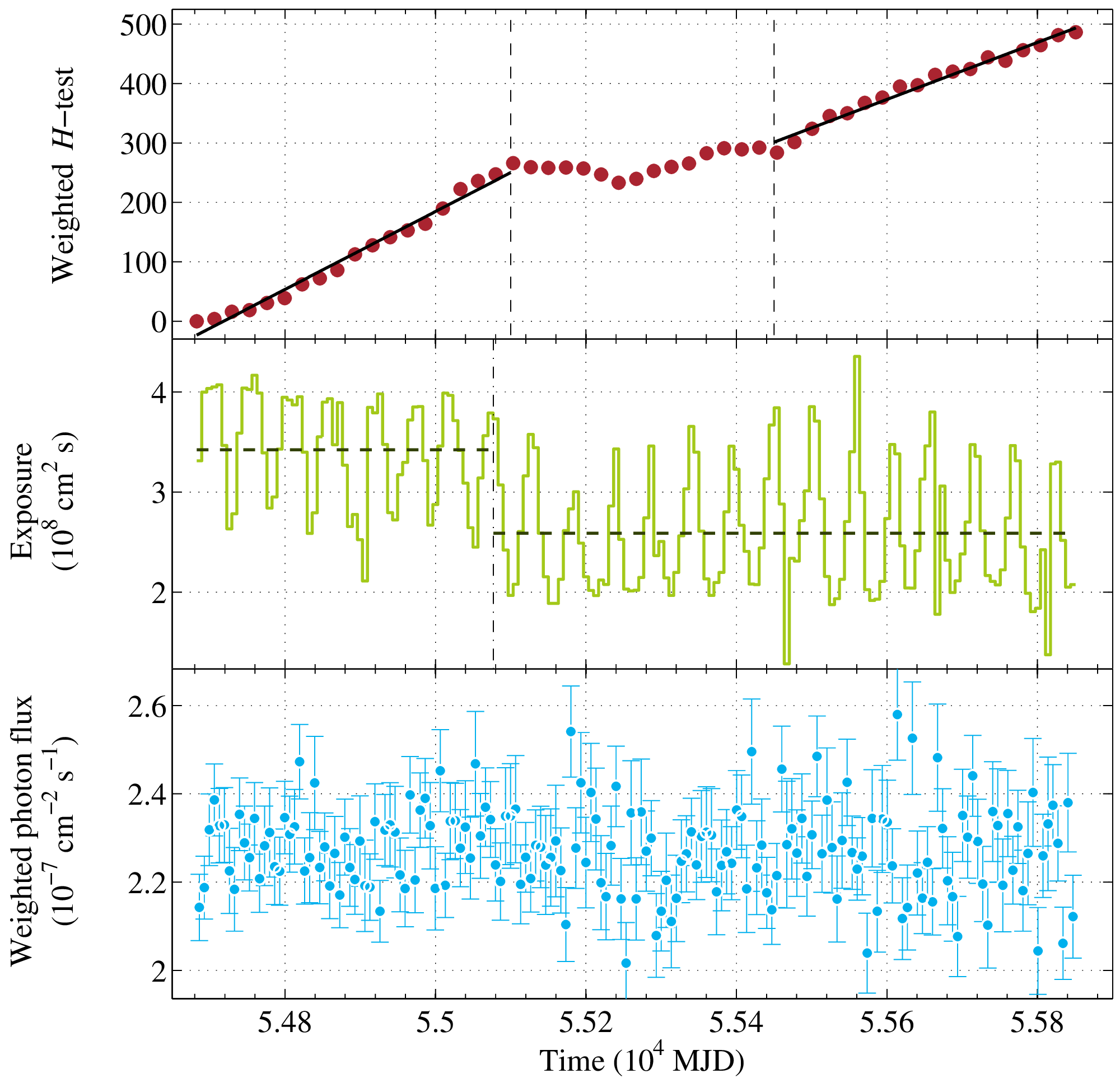}
\hfill
}
\caption{\label{f:wHtest} 
Upper panel: Weighted $H$-test statistic as a function of time
using the timing solution of Table~\ref{t:parameters} (dotted curve). 
Black solid lines represent separate linear fits before the 
nominal glitch (before MJD 55100), and after recovery of pulsations (from MJD 55450) 
where the measured slope is smaller by 27\%.
Middle panel: Exposure for the pulsar's sky location in time bins of size 6.7 days.
The average exposure (horizontal dashed line) after the LAT's observing-pattern
change at MJD 55077 (vertical dashed-dotted line) is smaller by 24\%. 
The exposure oscillation is at the 53.4-day precession period of the spacecraft orbit.
Lower panel: Sum of photon probability weights per time bin divided by exposure.
Binning in time is identical to the middle panel and
error bars represent 1$\sigma$ statistical uncertainties.\\
}
\end{figure}

The integrated pulse profile of \psr{} 
is fitted by two Lorentzian lines. The Full-Widths at Half Maxima (FWHM) 
for the first (P1) and second (P2) peak are $0.18 \pm 0.09$ and $0.13 \pm 0.05$,
respectively. The separation between the two is $\Delta = 0.24 \pm 0.04$. 
Figure~\ref{f:ph-vs-t-all} also exhibits a decrease in the ratio of peaks P1 and P2 with 
increasing energy. Also observed in other LAT pulsars 
\citep[see e.g.][]{FermiPSRCatalog,FermiJ2043+2740,FermiVela2}, this 
is thought to be caused by varying gamma-ray emission altitudes and curvature 
radii of the magnetic field lines as the pulsar rotates \citep{FermiVela2}.

Observing pulsed emission out to $\emax\sim 10$\,GeV implies
transparency to magnetic pair creation.  This bounds the altitude
$r$ of emission \citep[e.g.,][]{Abdo+2010-Crab}, giving
$r \geq (\emax B_{12} /2.7\,$GeV$)^{2/5} P^{-1/5} R_{\ast}$ in flat
spacetime (Story \& Baring, in preparation) for surface fields 
$10^{12} B_{12}$G. For \psr{} one obtains $r \gtrsim 6.1 R_{\ast}$,
precluding emission near the stellar surface.

Table~\ref{t:parameters} lists the best-fit values for $\Gamma$, $E_c$, and 
the photon and energy fluxes of \psr{} as derived from a spectral analysis of the 
region, restricted to photons with pulse phases between 0 and 0.5 
maximizing the pulsar's contribution. Further details on the spectral analysis of \psr{}
and a search for unpulsed gamma-ray emission from a putative pulsar wind nebula (PWN) 
in the off-pulse phase interval will be reported elsewhere \citep{Aharonian2012}.

\subsection{Pulsation Significance}

Over the 1168 days, the obtained timing solution yields a weighted $H$-test 
of~486.4, being extremely significant with a single-trial false alarm probability 
of~$\sim 10^{-92}$ \citep{KerrWeightedH2011}.

Figure~\ref{f:wHtest} investigates how this $H$-test value accumulates
over time. As expected, the signal increases linearly during times where 
the timing solution describes the pulsar's rotational behavior well (before the glitch 
at MJD 55100 and after MJD 55450). However, comparing the two different slopes from 
linear fits made during both time intervals, the slope during the post-glitch interval 
is smaller by about~27\%. 

This effect is traced back to a reduction in exposure resulting from a change 
in {\em Fermi}-LAT's sky-survey observing pattern on 2009 September 3  (MJD 55077).  
This change has decreased the average exposure at the pulsar's 
sky position by about 24\% (Figure~\ref{f:wHtest}, middle panel). 
Hence, the exposure reduction explains (within 3\%) the smaller rate in $H$-test increase 
over the later data (the $H$-test is linear in the number of photons). 
This is further confirmed by observing no significant variation in weighted photon flux 
over time (Figure~\ref{f:wHtest}, lower panel).

\section{Multiwavelength Observations}\label{s:mw}

\subsection{Radio and X-ray Counterpart Searches}\label{ss:radio-and-x-ray}

To determine whether \psr{} is also visible as a 
radio pulsar, a 1.7-hr observation has been conducted on 2011 May 27 
with the Green Bank Telescope (GBT) at 2~GHz. 
We also analyzed a 1-hr  archival GBT observation 
from on 2009 July 5 at 0.8~GHz \citep{Ransom2011}. 
Configuration details of both radio observations 
are found in Table~5 of \citet{Pletsch9pulsars}. 
Folding the data using the timing solution of Table~\ref{t:parameters}
and only searching in dispersion measure
revealed no radio pulsations. 

As in \citet{Pletsch9pulsars}, upper limits on the radio flux 
density are derived with the modified 
radiometer equation \citep{psrhandbook}.
The sensitivity loss due to the telescope pointing-direction offset 
from the pulsar position ($0.5\arcmin$ and $5.7\arcmin$, respectively)
is appropriately accounted for, analogously to \citet{Pletsch9pulsars}, leading to
radio flux upper limits of 9~$\mu$Jy for the 2~GHz observation, 
and 82~$\mu$Jy for the 0.8~GHz observation. These upper limits are comparable with 
those obtained for other LAT-discovered pulsars \citep{Ray2011,Pletsch9pulsars} 
suggesting that \psr{} is radio-quiet.

In searching for X-ray counterparts, two archival {\em Swift} observations 
(4.5~ks total exposure) cover the position of \psr{}, but no counterpart is observed. 
Using a 41.1-ks  {\em Suzaku} observation, we find a hint of X-ray emission at the 
pulsar location. However, the {\em Suzaku} large positional uncertainty ($\sim30\arcsec$)
and the low confidence ($\sim3\sigma$) of the counterpart candidate
preclude a firm X-ray detection of the pulsar.
The unabsorbed flux upper limit in the 0.3-10~keV energy range  
computed as in \citet{Marelli2011} is $5.0\times10^{-14}$ erg cm$^{-2}$s$^{-1}$,
similar to other LAT pulsars \citep{Marelli2011,Pletsch9pulsars}.

\subsection{A Likely Pulsar Wind Nebula Association}\label{ss:pwn}

The sky position of \psr{} coincides with the unidentified VHE source \tev{}
discovered during the H.E.S.S. Galactic Plane Survey \citep{Aharonian2008HESS}.
Although its extended TeV morphology suggests that \tev\, is composed of multiple sources,
limited exposure has hitherto prohibited statistically significant confirmation.
None of the previously known counterparts proposed in \cite{Aharonian2008HESS}
is capable of fully accounting for the VHE emission alone.

\psr{} is sufficiently energetic to power a PWN, further supporting this association.
With distance~$d$ to the pulsar, the spin-down flux at Earth is $\dot E/d^2$. 
If assuming a pseudo-distance\footnote{This  pseudo-distance is subject
to various caveats that translate to considerable uncertainties in this estimate 
of factors of a few \citep[][]{8gammapuls2010}.}  based on the observed
correlation between the gamma-ray luminosity ($0.1-100$\,GeV) and $\dot E$ 
\citep[following][]{8gammapuls2010} of $2\,\textrm{kpc}$ this
yields a large spin-down flux  
\mbox{$\dot E/d^2 \sim 1.5/d_{\mbox{\tiny 2 \textrm{kpc}}}^2 \times10^{36}$ erg s$^{-1}$ kpc$^{-2}$}
(where $d_{\mbox{\tiny 2 \textrm{kpc}}} = d/2$\,kpc),
which is comparable to similar systems \citep[][]{Carrigan2008}.

The integral energy flux  of \tev{} at the detector over the range $0.5-80$\,TeV is 
$G_{\rm VHE} \sim 5.8\times 10^{-11}$~erg cm$^{-2}$ s$^{-1}$ \citep{Sguera2009}.
Assuming isotropic emission this implies a conversion efficiency  
 $\eta = L_{\gamma} / \dot E = 0.5\, d_{\mbox{\tiny 2 \textrm{kpc}}}^2 \%$ ($0.5-80$\,TeV), 
similar to other suggested pulsar/PWN associations \citep{Hessels2008}.

\subsection{A Nearby Candidate Supernova Remnant}\label{ss:snr}

Because the pulsar is very young, we also consider the possible 
association with the nearby candidate supernova remnant (SNR)
G26.6$-$0.1.  This diffuse X-ray source 
was detected in the {\em ASCA} Galactic Plane Survey \citep{Bamba2003}. 
The best-fit absorption column density of the X-ray spectrum
\citet[][]{Bamba2003} yields a distance estimate of 1.3\,kpc.

The SNR candidate has an angular size of about $12\arcmin$ (FWHM), 
and its approximate geometric center lies $20\arcmin$ away from the 
pulsar's sky position. We note that the probability of chance superposition 
is also not negligible \citep[][]{Kaspi1998}.
To assess the SNR association, we assume the 1.3~kpc distance 
and an age of 5~kyr, and consider the corresponding pulsar transverse velocity
obtained as \mbox{$\sim1500$ $(d/1.3$~kpc $)$ km s$^{-1}$}. 
This is about 50\% faster than the highest velocity directly measured for 
a neutron star \citep{Chatterjee2005}.
However, the distance is quite uncertain and could 
be significantly smaller, which would reduce the resulting transverse velocity. 
Thus the SNR association appears unlikely, 
but cannot be fully excluded.

\section{Conclusions}\label{s:conclusions}

We report the discovery and follow-up study of \psr{} found in 
a new blind-search effort using {\em Fermi}-LAT data. 
The inferred parameters distinguish the pulsar as young and energetic. 
The characteristic age, $\tau = 4970\,\textrm{yr}$, makes
\psr{} the second-youngest pulsar ever found in a blind search of gamma-ray data, 
after PSR~J1023$-$5746 \citep{8gammapuls2010}, for which $\tau = 4610\,\textrm{yr}$.
Considering all rotation-powered pulsars known, \psr{} resides among 
the top 1\% of systems with largest~$\dot E$.
In deep radio observations no pulsations have been detected,
suggesting that \psr{} is radio-quiet, only detectable through 
its gamma-ray pulsations. 
Moreover, the pulsar experienced the largest glitch yet observed in 
any gamma-ray-only pulsar.

The pulsar's sky location coincides with the extended VHE source \tev. 
We have shown that the pulsar's energetics are likely to power
a PWN producing part of the H.E.S.S.-detected TeV emission.
This clearly motivates a further analysis of the region
at TeV energies to confirm such a PWN association.
With the LAT, an analysis for off-pulse emission of \psr{} will help studying
a potential PWN and the ambient medium \citep[e.g.,][]{FermiPWNsearch2011}.

\psr, the tenth object found in this new survey,
brings the number of gamma-ray pulsars undetected in radio to~33
\citep{Ray+2012}.
Together, these represent a significant increase
of the gamma-ray-only pulsar population and will be an important
contribution to the second {\em Fermi}-LAT Pulsar Catalog \citep{Fermi2PC}.
In addition to the improved search techniques \citep{Pletsch9pulsars},
the computational resources for the survey have recently been
significantly enhanced by using the volunteer computing 
system \EAH\footnote{http://einstein.phys.uwm.edu/}.
The unprecedented search sensitivity from the combination of these advances 
warrants optimism for further gamma-ray pulsar discoveries.

\acknowledgements

We thank M.-H.~Grondin, M.~Lemoine-Goumard and J.~M\'ehault
for helpful discussions on \tev.
This work was partly supported by the Max-Planck-Gesellschaft and by
U.S. National Science Foundation Grants 0970074 and 1104902.
The \emph{Fermi} LAT Collaboration acknowledges support from several agencies and institutes for both development and the operation of the LAT as well as scientific data analysis. These include NASA and DOE in the United States, CEA/Irfu and IN2P3/CNRS in France, ASI and INFN in Italy, MEXT, KEK, and JAXA in Japan, and the K.~A.~Wallenberg Foundation, the Swedish Research Council and the National Space Board in Sweden. Additional support from INAF in Italy and CNES in France for science analysis during the operations phase is also gratefully acknowledged.

\bibliographystyle{apj}

\end{document}